\documentclass[manuscript]{acmart}

\usepackage{tabularx}
\usepackage{enumitem}
\usepackage{multirow}
\usepackage{subcaption}
\usepackage{soul}
\usepackage{colortbl}

\AtBeginDocument{%
  }

\copyrightyear{2025}
\acmYear{2025}
\setcopyright{none}
\acmConference[CHI EA '25]{Extended Abstracts of the CHI Conference on Human Factors in Computing Systems}{April 26-May 1, 2025}{Yokohama, Japan}
\acmBooktitle{Extended Abstracts of the CHI Conference on Human Factors in Computing Systems (CHI EA '25), April 26-May 1, 2025, Yokohama, Japan}\acmDOI{10.1145/3706599.3706670}
\acmISBN{979-8-4007-1395-8/25/04}

\begin{document}

\title[Examining the Use and Impact of an AI Code Assistant on Dev. Productivity and Experience in the Enterprise]{Examining the Use and Impact of an AI Code Assistant on Developer Productivity and Experience in the Enterprise}

\author{Justin D. Weisz}
\email{jweisz@us.ibm.com}
\orcid{0000-0003-2228-2398}
\affiliation{
    \institution{IBM Research}
    \city{Yorktown Heights}
    \state{NY}
    \country{USA}
}

\author{Shraddha Kumar}
\email{shraddku@cisco.com}
\orcid{0000-0003-4997-2102}
\affiliation{
    \institution{Cisco Systems, Inc.}
    \city{Bangalore}
    \country{India}
}
\authornote{Work conducted while an employee of IBM Software, Kochi, India.}

\author{Michael Muller}
\email{michael_muller@us.ibm.com}
\orcid{0000-0001-7860-163X}
\affiliation{%
    \institution{IBM Research}
    \city{Cambridge}
    \state{MA}
    \country{USA}
}

\author{Karen-Ellen Browne}
\email{karen-ellen@ibm.com}
\orcid{0009-0004-1369-0300}
\affiliation{
    \institution{IBM Software}
    \city{Dublin}
    \country{Ireland}
}

\author{Arielle Goldberg}
\email{arielle.goldberg1@ibm.com}
\orcid{0009-0001-3868-2663}
\affiliation{%
    \institution{IBM Infrastructure}
    \city{Poughkeepsie}
    \state{NY}
    \country{USA}
}

\author{Ellice Heintze}
\email{ke.heintze@de.ibm.com}
\orcid{0009-0005-4834-769X}
\affiliation{%
    \institution{IBM Software}
    \city{Boeblingen}
    \country{Germany}
}

\author{Shagun Bajpai}
\email{Shagun.Bajpai@ibm.com}
\orcid{0009-0007-7413-9901}
\affiliation{%
    \institution{IBM Software}
    \city{Kochi}
    \country{India}
}

\renewcommand{\shortauthors}{Weisz et al.}


\begin{abstract}
    AI assistants are being created to help software engineers conduct a variety of coding-related tasks, such as writing, documenting, and testing code. We describe the use of the watsonx Code Assistant (WCA), an LLM-powered coding assistant deployed internally within IBM. Through surveys of two user cohorts (N=669) and unmoderated usability testing (N=15), we examined developers' experiences with WCA and its impact on their productivity. We learned about their motivations for using (or not using) WCA, we examined their expectations of its speed and quality, and we identified new considerations regarding ownership of and responsibility for generated code. Our case study characterizes the impact of an LLM-powered assistant on developers' perceptions of productivity and it shows that although such tools do often provide net productivity increases, these benefits may not always be experienced by all users.
\end{abstract}

\ccsdesc[500]{Human-centered computing~Empirical studies in HCI}
\ccsdesc[300]{Human-centered computing~Field studies}
\ccsdesc[500]{Software and its engineering~Collaboration in software development}
\ccsdesc[300]{Software and its engineering~Automatic programming}

\keywords{Generative AI, LLM, software engineering, productivity, code assistant}



\maketitle

\section{Introduction}

AI assistants powered by large language models (LLMs) are becoming increasingly prevalent in the workplace. A number of commercial and open-source coding assistants have been released for software engineers, developers, and data scientists\footnote{In this paper, we use the term ``developer'' as a catch-all to cover individuals who perform code-related work, including software engineers, architects, and data scientists.} who perform code-related work. These tools include GitHub Copilot\footnote{GitHub Copilot: \url{https://github.com/features/copilot}}, Amazon Q Developer\footnote{Amazon Q Developer: \url{https://aws.amazon.com/q/developer/}}, Gemini Code Assist\footnote{Gemini Code Assist: \url{https://cloud.google.com/gemini/docs/codeassist/overview}}, and IBM's watsonx Code Assistant\footnote{watsonx Code Assistant: \url{https://www.ibm.com/products/watsonx-code-assistant}}.

With this rapid proliferation of AI code assistants, it is important to understand their impact on developer productivity. We were provided with the opportunity to examine a new AI assistant under development within IBM in 2024 that supported general developer needs in common programming languages (e.g. Python, Java, JavaScript, C++, and more). Our team -- spanning product management, design, and research -- used a mixed-methods approach to characterize the assistant's impact on productivity.

In this paper, we report selected results from two studies: a large-scale survey of WCA users and small-scale usability testing. We discovered several interesting insights regarding the use of the assistant, the content it generated, and its impact on productivity and the developer profession. Our paper makes the following contributions to the CHI community:

\begin{itemize}
    \item We characterize the user experience of an LLM-based programming assistant under development within a large technology company. Our work examines the impact of the assistant on attitudinal measures of perceptions of productivity, complementing prior work that examined behavioral productivity metrics (e.g.~\cite{ziegler2022productivity, ziegler2024measuring}). We find that although the assistant did increase net productivity despite variability in the quality of its outputs, those gains were not evenly distributed across all users.
    \item We observed that \emph{understanding code} was the top use case, followed by the \emph{production of code}, indicating a need for further research in how AI code assistants can aid sensemaking tasks in code repositories.
    \item We identify a shared responsibility between people and AI systems in mitigating the risks of generated outputs.
\end{itemize}

\section{Related Work}

We outline three areas relevant to our study of AI code assistants: code-fluent LLMs and their incorporation into the software engineering workflow; the multi-faceted nature of productivity in software engineering; and studies of AI code assistants.

\subsection{Code-fluent LLMs and software engineering assistants}

Large language models that have been exposed to source code in their pre-training have demonstrated a high degree of aptitude in performing a variety of tasks: converting natural language to code (e.g.~\cite{xu2022ide, ahmad2021unified, feng2020codebert, guo2020graphcodebert}), converting code to natural language documentation (e.g.~\cite{luo2024repoagent, feng2020codebert}) or explanations (e.g.~\cite{nam2024using}), and converting code to code, such as by translating it from one language to another (e.g.~\cite{yang2024exploring, roziere2020unsupervised, ahmad2021unified, guo2020graphcodebert}) or by creating unit tests (e.g.~\cite{schafer2023empirical}). The introduction of the Codex model~\cite{chen2021evaluating} and its corresponding incorporation into software developers' IDEs through GitHub Copilot\footnote{GitHub Copilot: \url{https://github.com/features/copilot}} demonstrated how code-fluent LLMs could revolutionize the software development workflow. New agentic design patterns are enabling coding assistants to perform even more complex tasks, such as converting issues into pull requests~\cite{jimenez2023swe} and new feature descriptions into specifications and implementation plans\footnote{GitHub Copilot Workspace: \url{https://githubnext.com/projects/copilot-workspace}}.

\subsection{Software engineering productivity}

Productivity in software engineering is a complex, multi-faceted construct~\cite{sadowski2019rethinking, forsgren2021space}. It is often assessed via objective metrics of productivity that capture the ratio of output to effort~\cite{sadowski2019rethinking} (e.g. lines of code over time~\cite{devanbu1996analytical}, function points~\cite{low1990function}), the complexity of the software system~\cite{halstead1977elements, mccabe1976complexity}, or the presence of errors or defects~\cite{ko2005framework}. However, \citet{meyer2014software} considers ``when software developers perceive themselves to be productive and... unproductive''~\cite[p.1]{meyer2014software} as an important aspect of productivity.

\citet{cheng2022improves} outline a number of subjective and objective factors that impact developer productivity, including code quality, technical debt, infrastructure tools and support, team communication, and organizational changes and processes. In addition, researchers have found correlations between subjective and objective productivity metrics, such as the acceptance rate of suggested code~\cite{ziegler2024measuring} and the number of source code files owned by a developer~\cite{oliveira2020code} being correlated with perceived productivity.

The comprehensive landscape of software engineering productivity is captured by the SPACE framework~\cite{forsgren2021space}, which outlines both objective and subjective metrics across individuals, teams, and organizations. In this paper, we focus on attitudinal and human-centered  measures of productivity such as self-efficacy~\cite{ross2023programmer} and the impact of AI on the work process~\cite{weisz2022better}.

\subsection{Impact of LLM-based assistants on developer productivity}

Many studies have been conducted to examine the impact of LLM-based coding assistants on various aspects of productivity, albeit with mixed results~\cite{kuttal2021trade, vaithilingam2022expectation, ziegler2022productivity, ziegler2024measuring, wermelinger2023using, dakhel2023github, zhang2023practices, nguyen2022empirical, imai2022github, yetistiren2022assessing, pandey2024transforming, weisz2022better, ross2023programmer}. One early study by \citet{weisz2022better} examined AI-assisted code translation and found a net benefit to working with AI, though that benefit was not equally experienced by all participants. \citet{kuttal2021trade} examined human-human and human-AI pair-programming teams but did not find strong differences in outcomes such as productivity, code quality, or self-efficacy. 

\citeauthor{ziegler2024measuring} examined the impact of GitHub Copilot on developer productivity~\cite{ziegler2022productivity, ziegler2024measuring} and found that developers who used the tool self-reported higher levels of productivity. Contrarily, studies by \citet{imai2022github} and \citet{gitclear2024coding} both suggest that the quality of the code produced by GitHub Copilot may be harming productivity due to the number of lines that must be changed or deleted.

Another consideration for AI code assistants is their impact on the work process. Both \citet{barke2023grounded} and \citet{liang2024large} identified two complementary types of usage of GitHub Copilot: ``acceleration mode'' in which the tool aided developers when they knew what to do next, and ``exploration mode'' to help developers brainstorm potential solutions to coding problems when they were unsure of how to proceed.

\section{Case Study of an AI Code Assistant}

IBM's watsonx Code Assistant (WCA) is family of software engineering assistants that supports enterprise-specific use cases including IT automation\footnote{IBM watsonx\textsuperscript{\texttrademark} Code Assistant for Red Hat\textsuperscript{\textregistered} Ansible\textsuperscript{\textregistered} Lightspeed: \url{https://www.ibm.com/products/watsonx-code-assistant-ansible-lightspeed}} and mainframe application modernization\footnote{IBM watsonx\textsuperscript{\texttrademark} Code Assistant for Z: \url{https://www.ibm.com/products/watsonx-code-assistant-z}}. In mid-2024, a new variant of WCA, known as ``WCA@IBM,'' was released internally within IBM and was rapidly adopted by over 12,000 IBM developers. This variant provided general programming assistance in languages including Python, Java, JavaScript, C++, and more. It was implemented as plugins to VSCode and Eclipse, and it supported code generation from natural language, code autocompletion, code explanation and documentation, unit test generation, and conversational Q\&A. 

We had the opportunity to study internal usage of this new WCA\footnote{For simplicity, we refer to the internal deployment of ``WCA@IBM'' within IBM as ``WCA'' in this paper and we note that our use of ``WCA'' is not intended to refer to other products in the watsonx Code Assistant family.} variant from a subjective standpoint: how did it impact developers' perceptions of their own productivity? To address this question, we used a two-pronged approach: we conducted a survey of internal WCA users (N=669 respondents) and we conducted unmoderated usability testing in which participants used multiple WCA features to complete small programming tasks (N=15 participants).

Due to privacy concerns and sensitivities in data collection, we were only able to analyze attitudinal data in our studies. We did not have access to product telemetry or other behavioral data regarding use of WCA, such as the number of queries made throughout the study period, prompts sent to the model, or outputs generated by the model.

\subsection{Survey}

We developed a comprehensive survey to capture a wide range of data regarding the user experience of WCA and its impact on developer productivity. It was developed to meet the needs of multiple stakeholder groups, including product managers, designers, and researchers. We outline the different categories of measures in Table~\ref{tab:survey-measures}. We used a number of measures to capture different aspects of the WCA user experience, which we detail in Appendix~\ref{appendix:survey-instrument}.

Given the large number of measures included in the survey, we split it into two modules to prioritize the most important questions and help respondents avoid survey fatigue. Module 1 focused on understanding usage, usability, and demographics, and Module 2 focused more in-depth on motivations for use, trust, and content ownership \& responsibility. Each module took approximately 15-20 minutes to complete. Between Modules 1 and 2, respondents were asked whether they wanted to end the survey or spend additional time to provide more in-depth feedback.

\subsubsection{Recruitment \& participants}
\label{sec:survey-recruitment}

We launched a first round of data collection (Cohort 1) in May 2024 and received 105 responses. We launched a second round of data collection (Cohort 2) in July 2024, targeted to a group of 564 developers participating in a WCA training program. As completing the survey was part of the program, we eliminated the optional nature of Module 2 for this cohort.

A total of 669 WCA users responded to our survey. Despite the optional nature of Module 2 for Cohort 1, only 41 respondents did not complete this module (39.0\% of Cohort 1; 6.1\% of the total sample). Because we do not hypothesize any differences between the cohorts, we analyze them as a single group\footnote{We note that, despite the gap in time between the two cohorts, no major product enhancements beyond minor bug fixes were deployed; thus, the user experience between the two cohorts was equivalent.}.

Survey respondents held a wide range of developer roles, such as back-end developers (59.6\%), front-end developers (19.9\%), QA/test developers (19.3\%), and more. They held a wide range of tenures with IBM (< 6 months to 31+ years), years of experience working as a professional software engineer (0 to 10+ years), and hailed from a variety of geographies (primarily Americas and APAC). We show distributions of respondent demographics in Appendix~\ref{appendix:survey-respondents}.

\subsection{Unmoderated usability testing}

We conducted an unmoderated usability test to assess specific features of WCA. This test involved solving a small programming problem using different WCA features: code generation, chat, and code autocompletion. Participants then used WCA to generate explanations for their code and documentation in a README file. After each task, participants filled out a small questionnaire that asked about their experience. The usability test took approximately 40 minutes.

\subsubsection{Recruitment \& participants}
\label{sec:usability-recruitment}

We recruited 15 WCA users to complete the usability test based on the programming languages with which they worked and the frequency of their WCA usage. Participants came from a variety of product teams, spanning areas including firmware development, security, mobile apps, databases, and more. About 43\% of participants used WCA regularly in their work.

\subsection{Ethics statement}
Our research followed IBM's AI Ethics framework\footnote{IBM AI Ethics: \url{https://www.ibm.com/impact/ai-ethics}}. Our survey was approved by an internal review committee, subject to restrictions regarding the collection of demographic information. Specifically, we were unable to collect age or gender identity, we collected geography and job role using specified sets of options, and we did not collect any personally-identifiable information (PII). All survey responses were anonymous\footnote{For Cohort 2 respondents, we preserved the anonymity of their feedback by collecting their email address in a separate form after they completed the survey to mark their completion of the training course. In addition, it is possible that an individual responded to both surveys. However, given that the number of respondents in Cohort 2 was vastly greater than that of Cohort 1, we were not overly concerned with the possibility of capturing repeated measures from the same set of respondents.}.

\section{Results}

Our data consists of a mix of quantitative measures from the survey and qualitative feedback from the survey and the unmoderated usability testing sessions. We generally report descriptive statistics on the quantitative measures as our research questions did not require making statistical comparisons. To analyze qualitative data, we conducted a reflexive thematic analysis~\cite{clarke2017thematic}, using a deductive approach as our product management and design stakeholders had clear priorities for which aspects of the user experience on which to focus. We summarize our key results in Table~\ref{tab:survey-measures}.
    
In presenting our results, we refer to survey respondents as \texttt{Rx.yyy}, where \texttt{x} corresponds to their cohort (1 or 2) and \texttt{yyy} is a unique identifier. We refer to participants in the unmoderated usability study as \texttt{Pxx}, where \texttt{xx} is a unique identifier. Collectively, we refer to both groups as ``WCA users.''

\begin{table*}[htp]
    \centering
    \begin{tabularx}{\linewidth}{>{\raggedright}p{2.5cm}XXl}
        \toprule
        \textbf{Category} & \textbf{Description} & \textbf{Summary of Findings}  & \textbf{Section} \\
        \midrule
        Motivations, use, and non-use & Why and how WCA was used or not used
                                      & Top use cases focused on code understanding; ``off-label'' usage by content designers; unmet needs for specialized technologies (e.g. DB2, Maximo)
                                      & \ref{sec:motivations-and-use}
                                      \\
        \midrule
        Use of generated content      & Ways that generated content was reviewed \& used 
                                      & Content modified before use; outputs also used for learning and inspiration
                                      & \ref{sec:use-of-generated-content} \\
        \midrule
        Impact on\allowbreak productivity & Impact of WCA on various dimensions (effort, speed, work quality, self-efficacy) 
                                          & Small net productivity improvement, but with mixed and disparate impact
                                          & \ref{sec:impact-on-productivity} \\
        \midrule
        Authorship \&\allowbreak responsibility   & Who deserves authorship credit and who is responsible for avoiding inclusion of copyrighted IP? 
                                      & WCA deserving of authorship credit for co-creative activity; users and WCA have a joint responsibility to avoid inclusion of copyrighted IP
                                      & \ref{sec:authorship-and-responsibility} \\
        \midrule
        Impact on job role            & How AI assistants might change the developer profession 
                                      & AI lets developers focus on higher-level tasks; potential for deskilling; increased productivity translates into increased expectations
                                      & \ref{sec:impact-on-job-role} \\
        \bottomrule
    \end{tabularx}
    \caption{Categories of measures included in the user experience survey and a summary of our findings. We provide a complete listing of all survey questions in Appendix~\ref{appendix:survey-instrument}.}
    \Description{Categories of measures included in the user experience survey and a summary of our findings. We provide a complete listing of all survey questions in Appendix~\ref{appendix:survey-instrument}.}
    \label{tab:survey-measures}
\end{table*}

\subsection{Motivations, use, and non-use}
\label{sec:motivations-and-use}

\citet{liang2024large} examined a number of motivations for developers to use (or not use) AI programming assistants. Compared to their sample of AI code assistant users \emph{in the wild}, our respondents were less interested in having a code autocompletion feature (47\% vs. 86\%) or finishing their programming tasks faster (59\% vs. 76\%). Rather, they were more interested in discovering potential ways or starting points to solve their programming challenges (63\% vs. 50\%). In addition, our respondents were interested in exploring new tools (68\%), they wanted to know how their work might change in the future (64\%), they felt a responsibility to try new IBM products (64\%), and their management recommended using WCA (60\%). We provide a more detailed comparison of motivations for use in Appendix~\ref{appendix:motivations-comparison}.

We asked respondents about the kinds of tasks they conducted with WCA within their prior two weeks of usage. Prior research on LLM-based code assistants suggests that the core use is to \emph{generate code}~\cite{liang2024large}. By contrast, the most popular use cases for WCA related to \emph{understanding} code, either via explanations (71.9\%) or by receiving answers to general programming questions (68.5\%). R1.92 summarized the utility of WCA for code understanding: \emph{``I use WCA for two main things: Explaining code other people wrote that I am working with for the first time. Explaining code I have written that isn't working as expected, because sometimes WCA can tell me why it isn't working and give tips on how to fix it. It can often catch typo-like errors that I overlooked.''} R1.45 pointed out how explanations saved them time: \emph{``I like its ability to explain functions of code which could take a bit to understand. It can save a lot of time.''} R2.177 described their use of explanations to \emph{``explore areas of code I'm not familiar with.''} Respondents did use WCA to generate code (55.6\%), documentation (39.6\%), and tests (35.7\%), but to a lesser extent. We detail additional purposes of use in Appendix~\ref{appendix:purposes-of-use}.

Some respondents used WCA to improve specific qualities or characteristics of code, including its readability (17.9\%), maintainability (15.2\%), performance (12.8\%), and security (6.7\%). Usability test participants also indicated the importance of qualities including readability, maintainability, reliability, scalability, and testability. These desires for AI assistance beyond mere code generation mirror user needs for controlling code-fluent LLM outputs discovered by \citet{houde2022user}.

A small number of respondents reported not making use of WCA within the prior two weeks (N=28; 4.2\%). They indicated that it was faster to do the work themselves (39.3\%), they didn't feel that WCA provided helpful suggestions (32.1\%), they hadn't conducted any code-related work within the prior two weeks (25.0\%), and WCA didn't support their functional or non-functional requirements (14.3\%). Multiple respondents indicated a need for WCA to handle the specialized technologies with which they worked, such as R2.279, who commented, \emph{``I am a db2 developer and WCA's knowledge on db2 internals is poor.''} R2.365 also noted,  \emph{``it does not seem a good tool to search about IBM Products/offerings... such as Maximo Application Suite and MAS Ansible DevOps automation.''} In enterprise scenarios, it may be even more important to support the specialized technologies used in these environments to increase the utility of such assistants.

One interesting theme that stood out to us was a reluctance to use WCA because it may reflect negatively upon the user. R1.49 explained, \emph{``I just don't use code generated by WCA@IBM... obviously I would not want to be seen with generated code in my PRs, so embarrassing!''} This sentiment was echoed by R1.92, who said, \emph{``I imagine some people could find it embarrassing because the tool is so new and very few people on my team use it. There is an inherent suspicion against AI-generated code.''} Although only two respondents raised this issue, it may be important for organizational leaders to foster a culture in which AI assistance is viewed as acceptable to garner wide-spread adoption.

Finally, we discovered a small group of technical writers who found utility in using WCA to understand technical concepts without having to disturb their developer counterparts. R1.57 remarked, \emph{``My favorite feature is to understand the technical terms and code provided by the Dev. This reduces the time in understanding the API terms and code rather than discussing with the Dev.''} Multiple respondents desired using WCA to \emph{``create customer facing documentation''} (R1.27) and \emph{``help me writ[e] drafts following IBM content guidelines.''} (R1.68). These ``off-label'' use cases by people in developer-adjacent roles surprised us and indicate the importance of taking a holistic view on the potential beneficiaries of AI code assistants.

\subsection{Use of generated content}
\label{sec:use-of-generated-content}

Respondents reported using generated outputs -- code, documentation, unit tests, and explanations -- in different ways. Overall, the use of generated outputs without modification was not common (2-4\% of respondents reported doing this, depending on output type); rather, respondents often modified outputs before using them (9-19\%) or used them for another purpose, such as learning something new (23-35\%) or getting new ideas (24-37\%). Users described how \emph{``the results give me new ideas''} (R2.180) and \emph{``It is very helpful to get started writing code in a new language''} (R2.626) by \emph{``recommend[ing] an approach I haven't thought of or I wasn't even aware of''} (R2.405). P1 described how \emph{``creating diagrams in markdown works from the code.''} Users also talked about how WCA helped them recall \emph{``concepts which may be I have forgotten during [the] course of time''} (R2.296) and aiding them when \emph{``[I] know what to do, but don't know how to do it or forgot about that.''} (R2.292). These uses reinforce the value that generative AI provides in helping people learn~\cite{yilmaz2023effect, becker2023generative, shetye2024evaluation, adeshola2023opportunities}.

\subsection{Impact on productivity}
\label{sec:impact-on-productivity}

We evaluated the impact of WCA on respondents' perceptions of productivity using multiple measures: 7-point semantic differential scales\footnote{These scales were coded from [-3, +3] such that ratings > 0 correspond to beneficial impact, ratings of 0 indicate no impact, and ratings < 0 correspond to detrimental impact.} that assessed effort, quality, and speed~\cite{weisz2022better} and a 4-item scale of self-efficacy\footnote{Items were rated on 5-point Likert scales. A confirmatory factor analysis indicated this scale was highly reliable (Cronbach's $\alpha = .91$).} (derived from~\cite{ross2023programmer}). Overall, respondents felt that WCA made their work easier (M (SD) = .78 (1.45), t(609) = 13.35, p < .001, 95\% CI = [.67, .90]), of a better quality (M (SD) = .66 (1.25), t(603) = 13.02, p < .001, 95\% CI = [.56, .76]), and faster (M (SD) = .57 (1.48), t(606) = 9.57, p < .001, 95\% CI = [.46, .69]), as indicated by means and 95\% CIs > 0. However, the magnitudes of productivity improvements were small, further evidenced by self-efficacy ratings falling around the scale midpoint (M (SD) = 3.20 (.93) of 5).

Despite net positive ratings, the benefits of WCA were not evenly distributed: 42.6\% of respondents felt that WCA made them less effective (self-efficacy $\leq 3$), whereas 57.4\% of respondents felt that WCA made them more effective (self-efficacy $> 3$). Users' comments shed light on WCA's mixed impact on productivity. Some users benefited from using WCA in \emph{acceleration mode}: \emph{``Its ability to suggest code and code autocompletion... improves my work productivity significantly.''} (R2.608). P11 remarked, \emph{``It saves time compared to write from scratch by myself.''} WCA helped R2.236 come up to speed faster in a new project: \emph{``I used WCA to help document and explain me new functionality in different classes to help understand it quicker, thus making my productivity faster.''} Other users were helped in \emph{exploration mode} when WCA \emph{``provid[es] different approaches towards the problem''} (P11), which R2.311 described as \emph{``my favourite feature.''} P6 also felt, \emph{``I like the code generation aspect... it helps me think about the solution.''}

Contrarily, imperfections in the quality of WCA's output may require additional user effort to identify and correct. R1.25 pointed out that, \emph{``Sometimes WCA goes off topic and it ends up wasting time that could've been used to finish up the work. If multiple retries are needed to get the desired result, it becomes counter-productive.''} (R1.25). P5 noted that \emph{``It hallucinated''} during the usability test. R2.658 highlighted the relationship between correctness and trust, saying, \emph{``If it doesn't have close to 100\% correctness, then I cannot trust its answers on topics that I don't actually know myself. It is limited to helping me speed up more routine tasks that I can do myself.''} R2.603 commented, \emph{``You have to spend time to verify it.''} P13 similarly felt, \emph{``WCA created code need[s] to be tested,''} and R2.183 lamented, \emph{``it is a burden to have to double check answers...I still don't have enough confidence to blindly trust the responses.''}

We assessed the quality of WCA's outputs on a 5-point scale\footnote{Quality was assessed separately for each type of generated content: code, documentation, unit tests, explanations, and Q\&A responses. These items formed a scale with high reliability (Cronbach's $\alpha = 0.83$ and thus were averaged into a single metric of quality.}: Very poor (1), Poor, Acceptable, Good, Very good (5). Respondents rated WCA's quality in the middle of the scale (M (SD) = 3.20 (.77)), indicating an ``acceptable'' quality but with room for improvement. This level of quality was good enough for some respondents, but not others:

\begin{quote}
    \emph{``I'm really impressed with the quality of simple python programs that can be generated.''} (R2.143)
\end{quote}
\begin{quote}
    \emph{``The code generated is of poor quality and often does not meet the stated requirements.''} (R1.62)
\end{quote}

Another factor that impacted perceptions of productivity was the speed of WCA's responses. We assessed speed on a 5-point scale: Very slow (1), Slow, Acceptable, Fast, Very fast (5). Speed was rated slightly below the scale midpoint (M (SD) = 2.88 (0.86)) and many respondents commented on how it needed to be improved: \emph{``the code suggestion needs to be faster. At the moment it does not keep up with my typing.''} (R1.61). 

It is clear that WCA had an impact on productivity, but its impact was mixed and disparate. Several respondents pointed out how they viewed the role of WCA as an ``intern'' or ``junior developer'' given its current performance:

\begin{quote}
    \emph{``In some ways I feel it's like an intern that just started on the project and does training on the job. Needs a lot of supervision, but can be handed some set of tasks.''} (R1.86)
\end{quote}
\begin{quote}
    \emph{``I tend to view it as a junior developer helping me out. It can generate code much more quickly than a junior developer, and usually of higher quality, but there are still often mistakes, edge cases, etc. that need to be fixed. So I always need to be mindful that the generated code has to be carefully reviewed and often manually updated before it can be used.''} (R1.11)
\end{quote}

\subsection{Authorship \& responsibility}
\label{sec:authorship-and-responsibility}

Working with WCA is a co-creative process~\cite{wu2021ai, rezwana2023user, moruzzi2024user} in which both human users and WCA are capable of shaping an artifact-under-production, source code. Given the novelty of this form of interaction, we were interested in understanding how WCA impacted developers' perceptions of their authorship\footnote{Authorship and ownership are related, but distinct concepts. Ownership provides certain rights over the work, such as holding the copyright; as our users are employed by a corporation that owns the intellectual property produced by their employees, we assessed feelings of authorship rather than ownership.} over code co-produced with WCA.

We considered how respondents felt about who authored code in four different scenarios: (1) reviewing WCA-generated code but implementing the functionality themselves; (2) pasting in WCA-generated code verbatim; (3) pasting in WCA-generated code but then modifying it; and (4) implementing an idea suggested by WCA.

Many respondents viewed themselves as the sole author when implementing functionality themselves (scenario 1: 57.5\%). Similarly, many respondents viewed WCA as the sole author when they pasted its outputs verbatim (scenario 2: 53.7\%). Interestingly, some respondents felt a \emph{shared sense of authorship} across all scenarios: when it produced code that they rewrote (scenario 1: 30.4\%), when they used its outputs verbatim (scenario 2: 27.8\%), when each party contributed to the code (scenario 3: 64.4\%), and even when WCA only contributed ideas (scenario 4: 39.8\%). These results suggest that new mechanisms may be needed to track co-creative activities and ensure that each party's contributions -- human and AI -- are properly attributed in ways that do not \emph{``steal the recognition and hardwork of the developers''} (R2.632).

Another aspect of code authorship concerns the responsibility for mitigating the risks of generated outputs~\cite{ibm2024airisk}, such as ensuring generated code does not violate the IP rights of other parties. Given the current legal climate concerning the reproduction of copyrighted works by LLMs (e.g.~\cite{butterick2022github, roth2024developers}), we also wanted to understand the extent to which our users felt responsible for ensuring that WCA did not reproduce copyrighted works\footnote{Our examination of this topic is not predicated on observations of WCA actually reproducing copyrighted works; \citet{mishra2024granite} provide an explanation of the data used to train the underlying Granite model.}.

\citet{liang2024large} found that 46\% of their participants had concerns that AI programming assistants may produce code that infringes on intellectual property. By contrast, most of our respondents (83.4\%) expressed similar concerns over WCA reproducing copyrighted materials not owned by IBM. In addition, most respondents felt that they had a responsibility (89.2\%) and that WCA had a responsibility (96.2\%) to ensure copyrighted materials were not included in their source code. R2.549 aptly summarized their own responsibilities: \emph{``...if I am the person who uses the tool, it's my responsibility know what is going to be part of the code or not.''} Conversely, R2.459 expected WCA to detect copyrighted material: \emph{``it seems like copyrighted material should be detected as copyrighted and if this is not something WCA can do, it should definitely immediately do [so].''} Many respondents expected that, \emph{``all content generated or reproduced by WCA@IBM adheres to copyright regulations,''} (R2.652) and \emph{``WCA needs to fully own responsibility for not reproducing copyrighted code''} (R1.86). Thus, protecting the integrity of a codebase is seen as a shared responsibility, and one for which developers may require new kinds of support.

\subsection{Impact on job role}
\label{sec:impact-on-job-role}

Respondents held a wide range of views on how WCA would change their job role, responsibilities, and purpose as a developer. Some users felt that there would be ``No Change'' (R1.13), or that \emph{``It does not [change my role]. Yet, anyway :o)''} (R1.52), because \emph{``I'm still better than WCA''} (R2.546) or \emph{``I see it as a tool to better my work experience''} (R2.142). Other users felt that their role would change, but weren't sure of how: \emph{``It confuses me on what my duty is as the boundary is not clear''} (R2.472). Echoing sentiments of WCA's role as a ``junior developer'' or ``intern,'' some users felt it would free them up to focus on the higher-level, creative aspects of their profession:

\begin{quote}
    \emph{``I believe WCA has large potential to generate boilerplate code and implement trivial pieces of code. This would save developers time and effort writing a lot of repetitive and mundane code. It will allow developers to focus on solving the harder, higher level, and more creative problems.''} (R1.24)
\end{quote}

R2.304 also pointed out that having an AI assistant focus on the \emph{``many dirty repeated tasks like security fixes, add unit test cases/ translate the unit test cases etc.''} would allow them to \emph{``save my time to focus on innovative feature design and development.''}

Users were sensitive to the potential deskilling aspects of AI assistance~\cite{rafner2022deskilling, wang2019human}. R2.185 espoused a view that, \emph{``[I] [d]on't really like to use AI, makes people lazy and promotes to concept of not to think.''} R2.547 expressed the same concern, albeit more bluntly: \emph{''I suspect we're all going to get a lot stupider, doing a worse job of maintaining larger amounts of worse code.''} However, technological acceptance takes time~\cite{davis1989technology}, and new technologies have a learning curve before people can use them effectively; for LLMs, effective use requires learning the art of prompt engineering~\cite{lo2023art, marvin2023prompt}. WCA is no exception: \emph{``Very few people on my team have tried it, and many aren't sure how to engineer prompts to get effective answers yet''} (R1.92).

One final impact we noticed on the role of the developer is a consequence of AI augmenting human abilities: when a developer is more productive with AI, the expectations of their productivity may also be increased. This sentiment was captured by P6, who said, \emph{``Since our team is expected to use WCA our management is expecting more work in one sprint as compared to before.''}

\section{Discussion \& Conclusions}

Our examination of WCA usage revealed a number of insights on the use of AI code assistants within the enterprise and their impact on developer productivity.

\begin{itemize}
    \item \textbf{Code understanding is a top use case.} We anticipated that code generation would be a top use case due to the nature of generative AI and previous examinations of AI code assistants~\cite{liang2024large}. Rather, WCA was primarily used to support code understanding, either via explanations of existing code or through answers to general questions. This observation motivates the need for further research into how AI code assistants can help developers conduct sensemaking tasks in existing code repositories and bolster their knowledge of programming languages and concepts.
    \item \textbf{Software engineering is a co-creative activity.} Users generally did not accept the code produced by WCA without first reviewing and modifying it, suggesting that overreliance problems may not be as prevalent with AI code assistants as in other domains~\cite{buccinca2021trust, ashktorab2021ai, ashktorab2024emerging}. Additionally, they felt a shared sense of authorship over the code produced with WCA, even in instances when WCA only provided ideas and not actual source code. This observation suggests that new mechanisms for tracking authorship attribution may be necessary to give each party proper credit for their work, in addition to tracking the provenance of contributions within a larger source code repository.
    \item \textbf{Perfection is not required.} Echoing the title of the paper by \citet{weisz2021perfection}, we similarly observed that (some) developers felt productivity improvements from WCA despite variability in its quality and speed. We anticipate that quality and speed improvements will help more developers receive these benefits. We also observed that some developers may require additional training in effectively prompting LLMs. 
    \item \textbf{Mitigating risks in generated outputs is a joint responsibility.} LLMs have the potential to reproduce materials contained within their training data~\cite{bender2021dangers}; in the domain of software engineering, there is a risk of contaminating a codebase with copyrighted material or code subject to specific licenses (e.g. GPL). Users felt that they themselves, as well as WCA, were responsible for minimizing this risk. New socio-technical approaches that combine algorithmic methods (e.g. code similarity detection~\cite{novak2019source, jain2024llm}) with intelligent, human-driven review interfaces, may be needed to minimize these risks.
    \item \textbf{Developers are worried about deskilling and early adoption.} Users were concerned that their use of AI assistants may result in a loss of skills. In addition, two users were reluctant to use WCA due to perceived negative social consequences of being an early adopter. These findings indicate a need to more clearly articulate the benefits of AI code assistants and an organization's stance on their use: the developer profession will change to focus on higher-level, creative aspects of the work, and clear organizational policies can help employees feel more comfortable using AI code assistants in their work.
\end{itemize}

Our case study represents the use of a specific AI code assistant at a specific point of time (the summer of 2024). With the rapid pace of advancement in AI, we anticipate some issues experienced by our users, especially regarding quality and speed, will naturally diminish as model performance increases. What won't change is the need for human ingenuity and insight to determine \emph{what} software systems to build, even if the mechanics of \emph{how} those systems are built are increasingly aided by AI.


\begin{acks}
    We thank Katharina Schippert and Robin Auer for their support in defining the WCA user research program and recruiting participants for our studies. We also thank Keri Olson and Melissa Modjeski whose support made this research possible. Finally, we thank all of our users who provided us with valuable feedback.
\end{acks}

\bibliographystyle{ACM-Reference-Format}
\bibliography{references}

\newpage
\appendix
\section{Survey instrument}
\label{appendix:survey-instrument}

In this section, we provide a listing of the survey questions discussed in this case study. Some questions included in the survey have been omitted as their relevance to this case study is low, but their inclusion was important for other internal stakeholders.

\subsection{Usage within the past two weeks}

\textit{The survey began by asking whether the respondent had used WCA within the past two weeks.}

\begin{itemize}[leftmargin=0pt, itemindent=2em]
    \item[1.] Have you used WCA within the past 2 weeks?
    \item Yes
    \item No
\end{itemize}

\subsection{Motivations for use}

\textit{If the respondent indicated they used WCA within the past two weeks, they were asked about their motivations for using WCA.}

\begin{itemize}[leftmargin=0pt, itemindent=2em]
    \item[2a.] For what purposes did you use WCA? Please check all that apply. Please respond based only on your usage of WCA within the past 2 weeks.
    \item Generate code (or code snippets) in the chat
    \item Generate code via autocompletion in the source editor
    \item Generate documentation
    \item Generate tests
    \item Explain code
    \item Answer questions about APIs or libraries
    \item Answer general programming questions
    \item Generate alternate ways to implement a method or functionality
    \item Refactor code
    \item Optimize code for performance (e.g. runtime or memory usage)
    \item Translate code to another programming language
    \item Fix code
    \item Improve the readability of code
    \item Improve the maintainability of code
    \item Identify security issues in code
    \item Leave feedback for the WCA development team
\end{itemize}

\begin{itemize}[leftmargin=0pt, itemindent=2em]
    \item[2b.] For what other purposes did you use WCA? Please respond based only on your usage of WCA within the past 2 weeks.
    \item \emph{Open-ended response}
\end{itemize}

\subsection{Reasons for non-use}

\textit{If the respondent indicated they had not used WCA in the past two weeks, they were then asked about why. These items were based on \citet{liang2024large} and expanded upon by us.}

\begin{itemize}[leftmargin=0pt, itemindent=2em]
    \item[3a.] For what reasons have you not used WCA within the past 2 weeks? Please check all that apply.
    \item WCA writes code that doesn't meet functional or non-functional (e.g. security, performance) requirements that I need
    \item It's hard to control WCA to get code that I want
    \item I spent too much time debugging or modifying code written by WCA
    \item I don't think WCA provided helpful suggestions
    \item I don't want WCA to have access to my code
    \item I write and use code that WCA wasn't trained on which limits its ability to provide assistance
    \item I found WCA's suggestions too distracting
    \item I don't understand the code generated by WCA
    \item I don't understand the documentation generated by WCA
    \item I don't understand the explanations generated by WCA
    \item Code generated by WCA doesn't perform well enough for my needs
    \item It is faster to do the work myself
    \item I have not done any code-related tasks in the past 2 weeks
\end{itemize}

\begin{itemize}[leftmargin=0pt, itemindent=2em]
    \item[3b.] Did you have any other reasons for not using WCA? What were they?
    \item \emph{Open-ended response}
\end{itemize}

\subsection{Likes \& dislikes}

\begin{itemize}[leftmargin=0pt, itemindent=2em]
    \item[4a.] Please explain what you like about WCA. What is your favorite feature? Why?
    \item \emph{Open-ended response}
\end{itemize}

\begin{itemize}[leftmargin=0pt, itemindent=2em]
    \item[4b.] Please explain what you dislike about WCA@IBM. What is your least favorite feature? Why?
    \item \emph{Open-ended response}
\end{itemize}

\subsection{Self-efficacy}

\textit{The items in this scale were derived from \citet{ross2023programmer} and were expanded upon by us. Each item was rated on a 5-point Likert scale: Strongly disagree, Disagree, Neither disagree nor agree, Agree, Strongly agree.}

\begin{itemize}[leftmargin=0pt, itemindent=2em]
    \item[5.] How would you characterize your experience using WCA?
    \item WCA helps me write better code
    \item WCA helps me write code more quickly
    \item I feel more productive when using WCA
    \item I spend less time searching for information while using WCA
\end{itemize}

\subsection{Speed}

\begin{itemize}[leftmargin=0pt, itemindent=2em]
    \item[6.] How would you characterize the speed of receiving a chat response from WCA?
    \item I have not used the chat feature, Unusably slow, Very slow, Slow, Acceptable, Fast, Very fast
\end{itemize}

\subsection{Copyright}

\begin{itemize}[leftmargin=0pt, itemindent=2em]
    \item[7a.] How concerned are you that WCA may reproduce copyrighted materials not owned by IBM?
    \item Not concerned at all, Slightly concerned, Moderately concerned, Concerned, Very concerned
\end{itemize}

\noindent\textit{Respondents were asked to rate the degree of responsibility for each party on a 5-point scale: Not at all responsible, Slightly responsible, Moderately responsible, Responsible, Very responsible.}

\begin{itemize}[leftmargin=0pt, itemindent=2em]
    \item[7b.] When using WCA, to what extent are the following parties responsible for ensuring that copyrighted materials not owned by IBM are not included in IBM source code?
    \item Me
    \item WCA
\end{itemize}

\subsection{Additional improvements}

\begin{itemize}[leftmargin=0pt, itemindent=2em]
    \item[8.] What single aspect of WCA should be immediately improved? Please explain.
    \item \emph{Open-ended response}
\end{itemize}

\subsection{Demographics}

\begin{itemize}[leftmargin=0pt, itemindent=2em]
    \item[9a.] How many years of service do you have at IBM (based on date of hire)
    \item < 6 months, 6 months–1 year, 1-2 years, 3-5 years, 6-10 years, 11-15 years, 16-20 years, 21-25 years, 26-30 years, 31+ years
\end{itemize}

\begin{itemize}[leftmargin=0pt, itemindent=2em]
    \item[9b.] How many years have you held a role as a professional software engineer (at IBM or other companies)?
    \item 0-5 years, 6-10 years, 10+ years
\end{itemize}

\begin{itemize}[leftmargin=0pt, itemindent=2em]
    \item[9c.] What is your job role? Please check all that apply.
    \item Back-End Developer 
    \item Compiler Technology Developer 
    \item Developer Advocate 
    \item DevOps Developer 
    \item Firmware Developer 
    \item Front-End Developer 
    \item L3 Support Engineer 
    \item QA/Test Developer 
    \item Software Architect 
    \item Software Performance Analyst 
    \item Other: \emph{free text}
\end{itemize}

\subsection{Module 2}

\textit{At this point in the survey, respondents were asked whether they wished to continue to provide additional, in-depth feedback.}

\begin{itemize}[leftmargin=0pt, itemindent=2em]
    \item[10.] Would you like to provide additional feedback on WCA?
    \item Yes
    \item No
\end{itemize}

\subsection{In-depth quality}

\textit{We asked respondents to rate the quality of each type of generated output on a 5-point scale: Very poor, Poor, Acceptable, Good, Very good. These ratings were averaged to produce an overall quality score.}

\begin{itemize}[leftmargin=0pt, itemindent=2em]
    \item[11.] How would you characterize the quality of WCA's outputs?
    \item Generated code (within chat or the source editor)
    \item Generated unit tests
    \item Generated documentation
    \item Generated explanations of code
    \item Answers to general questions (within chat)
\end{itemize}

\subsection{In-depth motivations for use}

\textit{These motivations were derived from \citet{liang2024large} and were expanded upon by us. Each item was rated on a 5-point scale: Not important at all, Slightly important, Moderately important, Important, Very important.}

\begin{itemize}[leftmargin=0pt, itemindent=2em]
    \item[12a.] What motivations did you have for using WCA?
    \item To have an autocomplete or reduce the amount of keystrokes I make
    \item To finish my programming tasks faster
    \item To skip needing to go online to find specific code snippets, programming syntax, or API calls I'm aware of, but can't remember
    \item To discover potential ways or starting points to write a solution to a problem I'm facing
    \item To find an edge case for my code I haven't considered
    \item My line management recommended I use it
    \item My colleagues recommended I use it
    \item I like to explore new tools
    \item It is my responsibility to try new IBM products
    \item I wanted to know more about how my work might change in the future
\end{itemize}

\begin{itemize}[leftmargin=0pt, itemindent=2em]
    \item[12b.] Did you have any other reasons for using WCA? What were they?
    \item \emph{Open-ended response}
\end{itemize}

\subsection{Use of generated content}

\textit{If the respondent indicated they used WCA within the past two weeks, they were asked these questions about how they made use of content generated by WCA.}

\begin{itemize}[leftmargin=0pt, itemindent=2em]
    \item[13a.] Within the past 2 weeks, how have you used content produced by WCA? Please check all that apply. Please respond based only on your usage of WCA within the past 2 weeks.
    \item Code snippets (excluding unit tests)
    \item Unit tests
    \item Natural language explanations
    \item Documentation
\end{itemize}

\noindent\textit{The items above were rated on the following scale: I included it in my code without modification, I included it in my code and made changes to it, I used it to give me new ideas, I used it to learn something new, I didn't make use of it}

\begin{itemize}[leftmargin=0pt, itemindent=2em]
    \item[13b.] Within the past 2 weeks, how did you decide when to use WCA versus completing a task yourself? Please respond based only on your usage of WCA within the past 2 weeks.
    \item \emph{Open-ended response}
\end{itemize}

\subsection{Impact on productivity}

\textit{These scales were reproduced from \citet{weisz2022better}. Each item was rated on a 7-point semantic differential scale.}

\begin{itemize}[leftmargin=0pt, itemindent=2em]
    \item[14.] Do you feel that WCA makes your work...
    \item More difficult / Easier
    \item Slower / Faster
    \item Of the worst quality / Of the best quality
\end{itemize}

\subsection{In-depth use of generated content}

\textit{The items in this section were rated on a 5-point scale: Never, Rarely, Sometimes, Often, Always.}

\begin{itemize}[leftmargin=0pt, itemindent=2em]
    \item[15.] How often do you take these actions to evaluate content generated by WCA?
    \item Quickly check the generated code for specific keywords or logic structures
    \item Compile, type check, lint, and/or use an in-IDE syntax checker
    \item Execute the generated code
    \item Examine details of the generated code's logic in depth
    \item Consult API documentation
\end{itemize}

\subsection{Authorship}

\textit{These items in this question were rated on a 4-point scale: Me, Both of us, WCA, I'm unsure.}

\begin{itemize}[leftmargin=0pt, itemindent=2em]
    \item[16a.] When using WCA, who do you consider to be the author of the code when...
    \item I paste WCA-generated code into my source file verbatim
    \item I paste WCA-generated code into my source file but then make edits to it
    \item I review code generated by WCA, but ultimately implement the functionality myself
    \item I implement an idea suggested by WCA
\end{itemize}

\begin{itemize}[leftmargin=0pt, itemindent=2em]
    \item[16b.] How does WCA change your perceptions of your job role, responsibilities, and purpose as a developer?
    \item \emph{Open-ended response}
\end{itemize}

\subsection{Final comments}

\begin{itemize}[leftmargin=0pt, itemindent=2em]
    \item[17a.] Please describe your ideal vision for how WCA could help boost your productivity as a developer. Consider the following ideas: What additional features or capabilities does it need? How would it fit into your workflow? What kinds of tasks would you use it for and what kinds of tasks would you not use it for?
    \item \emph{Open-ended response}
\end{itemize}

\begin{itemize}[leftmargin=0pt, itemindent=2em]
    \item[17b.] Is there any other feedback you would like to provide about WCA or this survey?
    \item \emph{Open-ended response}
\end{itemize}

\section{Respondents \& participants}
\label{appendix:survey-respondents}

\begin{figure*}[ht]
    \centering
    \begin{subfigure}{0.33\textwidth}
        \includegraphics[width=\textwidth]{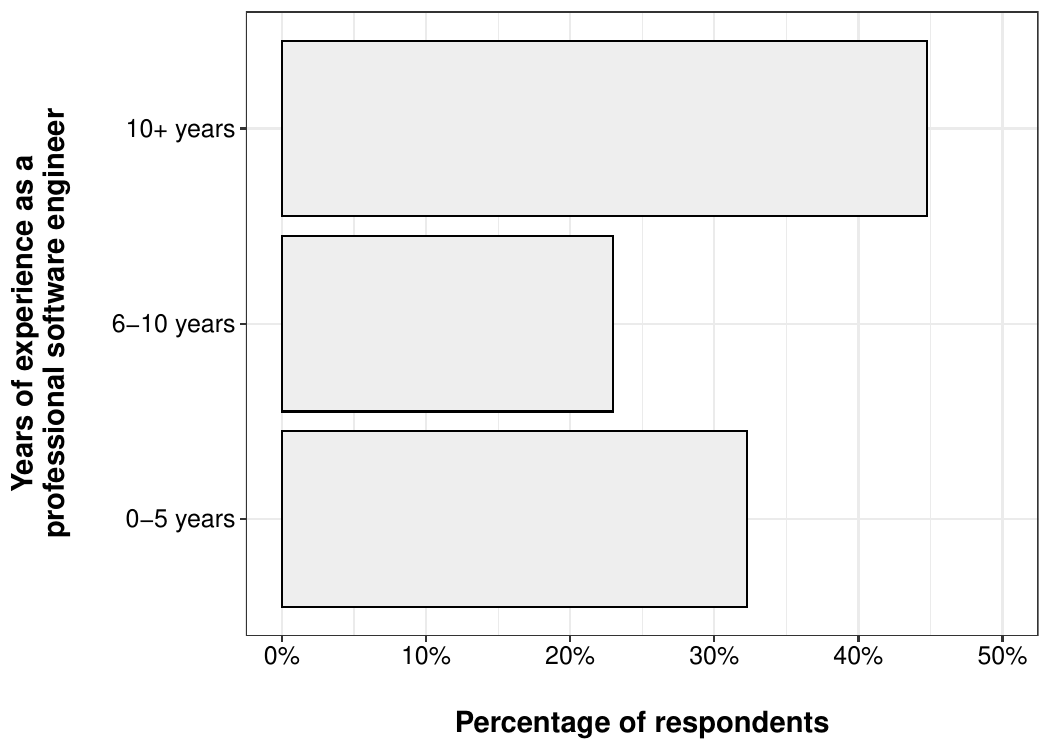}
        \caption{}
        \label{fig:survey-demographics-yoe}
    \end{subfigure}
    \begin{subfigure}{0.33\textwidth}
        \includegraphics[width=\textwidth]{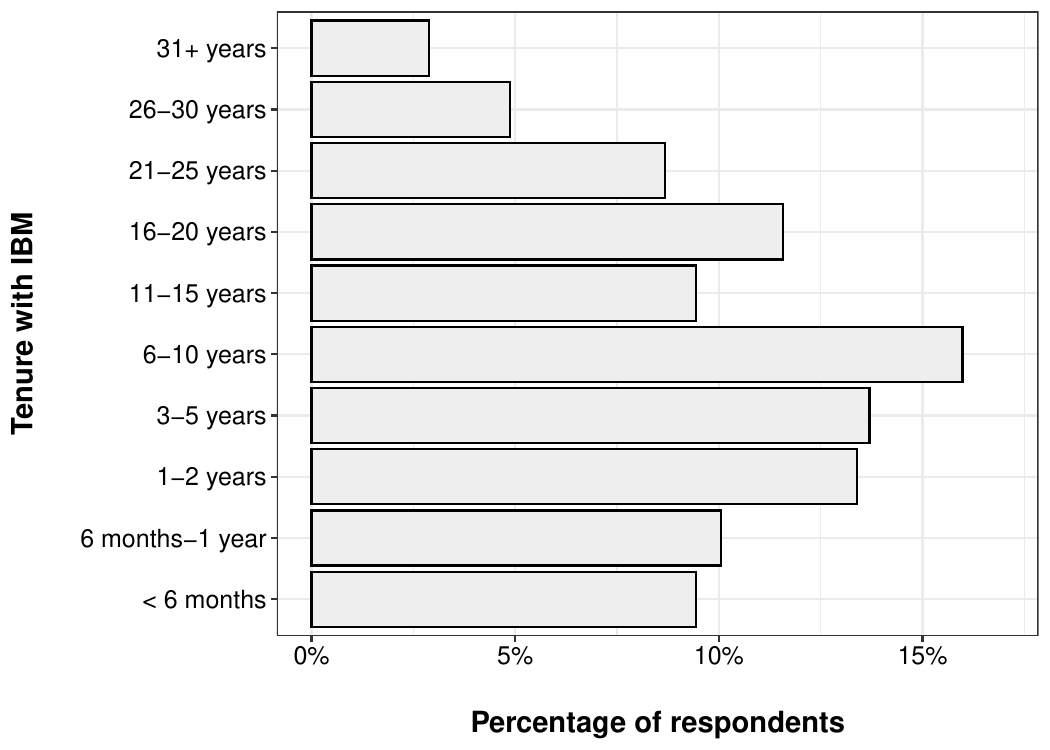}
        \caption{}
        \label{fig:survey-demographics-tenure}
    \end{subfigure}
    \begin{subfigure}{0.33\textwidth}
        \includegraphics[width=\textwidth]{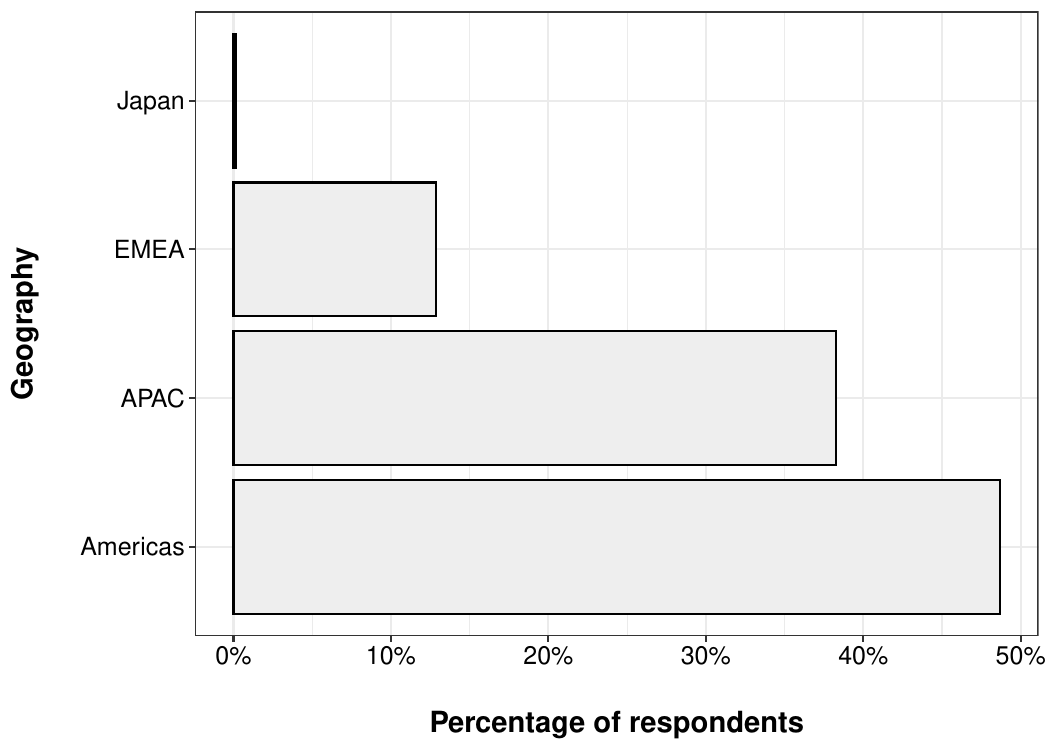}
        \caption{}
        \label{fig:survey-demographics-geo}
    \end{subfigure}
    \caption{Survey respondent demographics: (a) years of experience as a professional software engineer, (b) tenure with IBM, and (c) geography.}
    \Description{The figure consists of three horizontal bar graphs. The first graph (A) depicts the percentage of respondents and their years of experience as a professional software engineer. Years of experience includes three categories: 0 to 5 years (31.7\%), 6 to 10 years (22.6\%), and 10 plus years (44.0\%). The second graph (B) depicts the percentage of respondents and their years of tenure with IBM. Tenure with IBM includes ten categories; less than 6 months (9.3\%), 6 months to one year (9.9\%), 1 to 2 years (13.1\%), 3 to 5 years (13.4\%), 6 to 10 years (15.7\%), 11 to 15 years (9.3\%), 16 to 20 years (11.4\%), 21 to 25 years (8.5\%), 26 to 30 years (4.8\%), and 31 plus years (2.8\%). The third graph (C) depicts the percentage of respondents per geography. Geography includes four categories: Japan (less than 1\%), EMEA (12.6\%), APAC (37.4\%), and Americas (47.5\%).}
\end{figure*}

For survey respondents, we show the distributions of their years of experience as a professional software engineer in Figure~\ref{fig:survey-demographics-yoe}, tenure with IBM in Figure~\ref{fig:survey-demographics-tenure}, and geography in Figure~\ref{fig:survey-demographics-geo} (survey questions 9a-9c).

\section{Motivations for using AI programming assistants}
\label{appendix:motivations-comparison}

\newlength\BW\setlength\BW{0.76in}
\newlength\BH\setlength\BH{2ex}
\newcommand{\barRule}[2][gray]{\textcolor{#1}{\rule{#2\BW}{\BH}}}
\definecolor{VI}{HTML}{2F70CD}
\definecolor{I}{HTML}{75ACF0}
\definecolor{MI}{HTML}{DCDCDC}
\definecolor{SI}{HTML}{EFB76F}
\definecolor{NI}{HTML}{E78E35}
\newcommand*\BarStack[7]{#1~\barRule[VI]{#2}\barRule[I]{#3}\barRule[MI]{#4}\barRule[SI]{#5}\barRule[NI]{#6}~#7}
\newcommand*\BarChip[2]{\textcolor{#1}{\rule{2ex}{\BH}}~#2}

Table~\ref{tab:motivations-for-use} shows a detailed comparison of motivations for using or not using AI programming assistants between our survey respondents and the software engineers who responded to the survey reported by \citet{liang2024large} (survey question 12a). We include a number of additional motivations (A1-10) identified as important by our product management team. We also note that respondents were only asked to select which motivations explained their non-use of WCA rather than rating them on the 5-point scale of importance; we therefore only report the percentage of respondents who indicated each motivation for non-use. In addition, we did not include all motivations for non-use from \citet{liang2024large} as some were not relevant for our enterprise context.

\begin{table*}[ht]
    \smaller
    \centering
    \begin{tabularx}{\linewidth}{lXp{1.15in}p{1.25in}}
        \toprule
        \multicolumn{2}{l}{\textbf{Motivations}} & \textbf{\citet{liang2024large}} & \textbf{Our respondents} \\
        \midrule
        \multicolumn{4}{l}{\textit{Motivations for use (\citet{liang2024large})}} \\
        M1  & To have an autocomplete or reduce the amount of keystrokes I make 
            & \BarStack{86\%}{0.55}{0.31}{0.08}{0.04}{0.02}{6.2\%} 
            & \BarStack{47\%}{0.18}{0.29}{0.23}{0.18}{0.12}{30\%} \\
        M2  & To finish my programming tasks faster 
            & \BarStack{76\%}{0.43}{0.33}{0.12}{0.10}{0.02}{12\%} 
            & \BarStack{59\%}{0.26}{0.33}{0.20}{0.15}{0.06}{21\%} \\
        M3  & To skip needing to go online to find specific code snippets, programming syntax, or API calls I'm aware of, but can't remember 
            & \BarStack{68\%}{0.41}{0.28}{0.17}{0.10}{0.04}{14\%} 
            & \BarStack{64\%}{0.30}{0.34}{0.20}{0.11}{0.06}{17\%} \\
        M4  & To discover potential ways or starting points to write a solution to a problem I'm facing 
            & \BarStack{50\%}{0.24}{0.26}{0.26}{0.15}{0.09}{24\%} 
            & \BarStack{64\%}{0.30}{0.34}{0.20}{0.11}{0.06}{17\%} \\
        M5  & To find an edge case for my code I haven't considered 
            & \BarStack{36\%}{0.15}{0.22}{0.19}{0.25}{0.19}{44\%}
            & \BarStack{48\%}{0.18}{0.29}{0.24}{0.18}{0.10}{28\%} \\
        \midrule
        \multicolumn{4}{l}{\textit{Additional motivations for use}} \\
        A1  & My colleagues recommended I use it
            & \multicolumn{1}{c}{--}
            & \BarStack{30\%}{0.09}{0.20}{0.32}{0.21}{0.18}{39\%} \\
        A2  & My line management recommended I use it
            & \multicolumn{1}{c}{--}
            & \BarStack{60\%}{0.29}{0.31}{0.21}{0.11}{0.07}{19\%} \\
        A3  & I like to explore new tools
            & \multicolumn{1}{c}{--}
            & \BarStack{68\%}{0.34}{0.35}{0.17}{0.11}{0.04}{14\%} \\
        A4  & I wanted to know more about how my work might change in the future
            & \multicolumn{1}{c}{--}
            & \BarStack{64\%}{0.28}{0.36}{0.18}{0.12}{0.06}{18\%} \\
        A5  & It is my responsibility to try new IBM products
            & \multicolumn{1}{c}{--}
            & \BarStack{64\%}{0.27}{0.37}{0.18}{0.13}{0.05}{18\%} \\
        \midrule
        \multicolumn{2}{l}{\textit{Motivations for non-use (\citet{liang2024large})}} \\
        M6  & Code generation tools write code that doesn't meet functional or non-functional (e.g., security, performance) requirements that I need
            & \BarStack{54\%}{0.31}{0.24}{0.12}{0.18}{0.15}{34\%}
            & \multicolumn{1}{c}{14.3\%} \\
        M7  & It's hard to control code generation tools to get code that I want
            & \BarStack{48\%}{0.24}{0.24}{0.18}{0.20}{0.14}{36\%}
            & \multicolumn{1}{c}{14.3\%} \\
        M8  & I spend too much time debugging or modifying code written by code generation tools
            & \BarStack{38\%}{0.18}{0.20}{0.18}{0.16}{0.28}{45\%}
            & \multicolumn{1}{c}{14.3\%} \\
        M9  & I don't think code generation tools provide helpful suggestions
            & \BarStack{34\%}{0.14}{0.20}{0.22}{0.22}{0.22}{46\%}
            & \multicolumn{1}{c}{32.1\%} \\
        M10 & I don't want to use a tool that has access to my code
            & \BarStack{30\%}{0.24}{0.06}{0.20}{0.10}{0.40}{51\%}
            & \multicolumn{1}{c}{--} \\
        M11 & I write and use proprietary code that code generation tools haven't seen before and don't generate
            & \BarStack{28\%}{0.08}{0.20}{0.14}{0.22}{0.36}{59\%}
            & \multicolumn{1}{c}{--} \\
        M12 & To prevent potential intellectual property infringement
            & \BarStack{26\%}{0.13}{0.13}{0.09}{0.10}{0.55}{66\%}
            & \multicolumn{1}{c}{--} \\
        M13 & I find the tool's suggestions too distracting
            & \BarStack{26\%}{0.14}{0.12}{0.24}{0.22}{0.28}{51\%}
            & \multicolumn{1}{c}{--} \\
        M14 & I don't understand the code written by code generation tools
            & \BarStack{16\%}{0.02}{0.14}{0.08}{0.27}{0.49}{76\%}
            & \multicolumn{1}{c}{3.6\%} \\
        M15 & I don't want to use open-source code
            & \BarStack{10\%}{0.02}{0.08}{0.08}{0.13}{0.69}{89\%}
            & \multicolumn{1}{c}{--} \\
        \midrule
        \multicolumn{4}{l}{\textit{Additional motivations for non-use}} \\
        A6  & Code generated by WCA doesn't perform well enough for my needs
            & \multicolumn{1}{c}{--}
            & \multicolumn{1}{c}{17.9\%} \\
        A7  & I don't understand the explanations generated by WCA
            & \multicolumn{1}{c}{--}
            & \multicolumn{1}{c}{3.6\%} \\
        A8  & I have not done any code-related tasks in the past 2 weeks
            & \multicolumn{1}{c}{--} 
            & \multicolumn{1}{c}{25.0\%} \\
        A9  & I write and use code that WCA wasn't trained on which limits its ability to provide assistance
            & \multicolumn{1}{c}{--}
            & \multicolumn{1}{c}{10.7\%} \\
        A10 & It is faster to do the work myself
            & \multicolumn{1}{c}{--}
            & \multicolumn{1}{c}{39.3\%} \\
        \midrule
        \multicolumn{4}{c}{\BarChip{VI}{Very important}\hspace{1em}\BarChip{I}{Important}\hspace{1em}\BarChip{MI}{Moderately important}\hspace{1em}\BarChip{SI}{Slightly important}\hspace{1em}\BarChip{NI}{Not important at all}} \\
        \bottomrule
    \end{tabularx}
    \caption{Comparison of motivations for using (M1-5) and not using (M6-15) AI programming assistants with \citet{liang2024large}, along with additional motivations (A1-10) examined in our survey. Percentages on the left-hand side (right-hand side) of each bar indicate the proportion of respondents who rated each motivation as ``Very important'' or ``Important'' (``Slightly important'' or ``Not important at all'') on a 5-point scale. Only 28 of our 669 respondents (4.2\%) indicated non-use of WCA within the prior two weeks.}
    \Description{Comparison of motivations for using (M1-5) and not using (M6-15) AI programming assistants with Liang et al. (2024), along with additional motivations (A1-10) examined in our survey. Percentages on the left-hand side (right-hand side) of each bar indicate the proportion of respondents who rated each motivation as ``Very important'' or ``Important'' (``Slightly important'' or ``Not important at all'') on a 5-point scale. Only 28 of our 669 respondents (4.2\%) indicated non-use of WCA within the prior two weeks.}
    \label{tab:motivations-for-use}
\end{table*}

\section{Purposes of use of WCA}
\label{appendix:purposes-of-use}

Figure~\ref{fig:survey-purposes} shows the distribution of purposes of use of WCA (survey question 2a). These data were reported by respondents who indicated they had used WCA within the prior two weeks (N=638).

\section{Distributions of productivity measures}
\label{appendix:productivity}

\begin{figure*}[ht]
    \centering
    \begin{subfigure}{0.33\textwidth}
        \includegraphics[width=\textwidth]{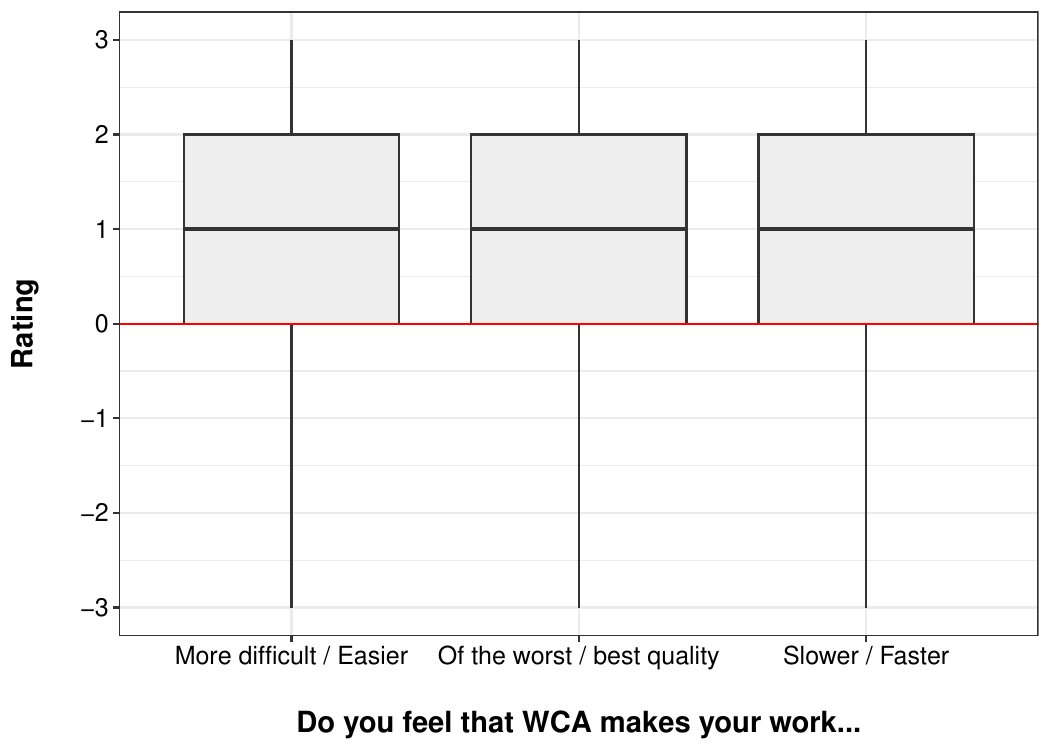}
        \caption{}
    \end{subfigure}
    \begin{subfigure}{0.33\textwidth}
        \includegraphics[width=\textwidth]{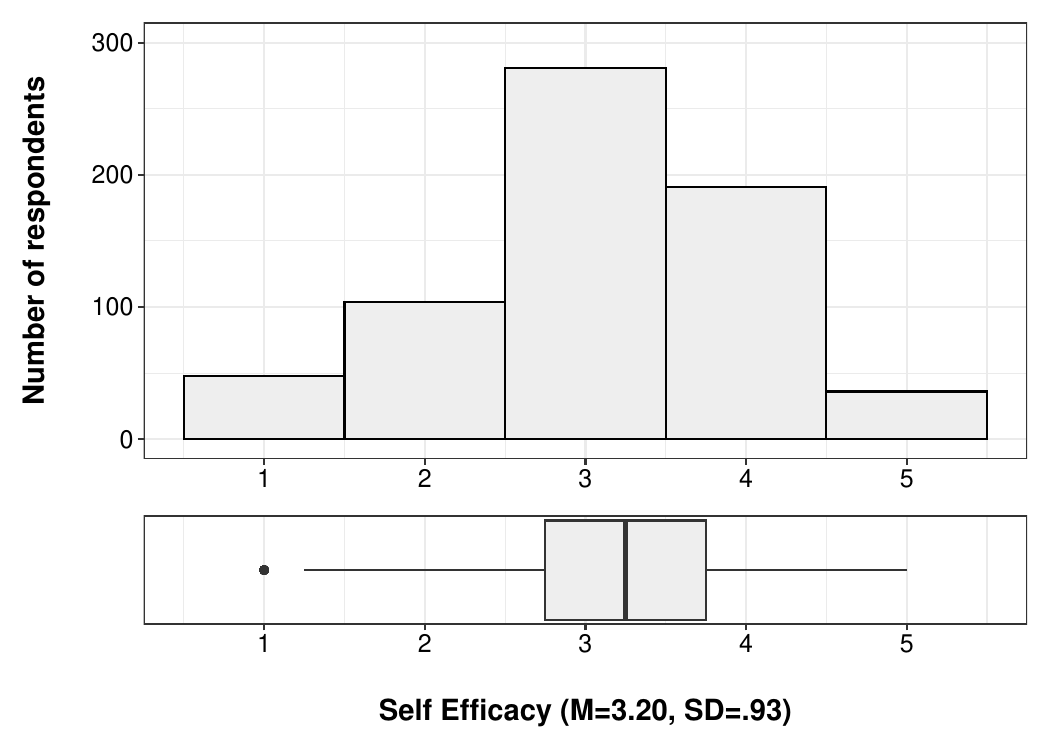}
        \caption{}
    \end{subfigure}
    \begin{subfigure}{0.33\textwidth}
        \includegraphics[width=\textwidth]{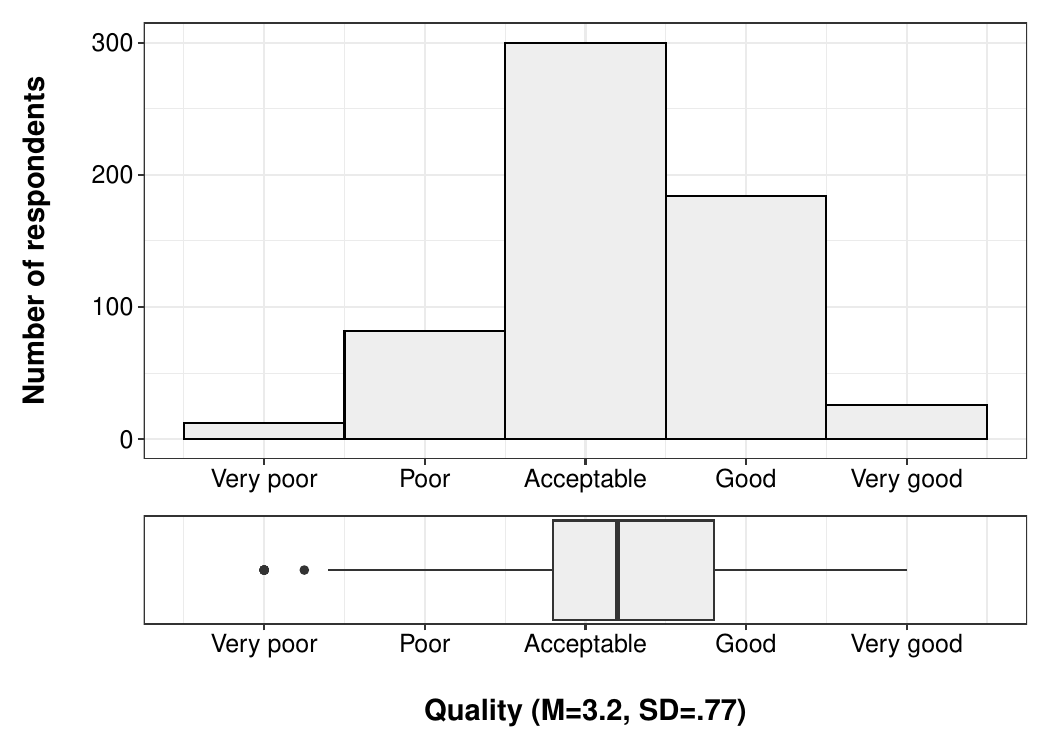}
        \caption{}
    \end{subfigure}
    \caption{Distributions of productivity and quality measures: (a) self-reported ratings of effort, quality of work, and speed on 7-point semantic differential scales (centered at 0), (b) distribution of self-efficacy scores, and (c) distribution of overall quality scores.}
    \Description{The figure consists of three graphs. The first graph (A) includes three box plots that depict how respondents feel that WCA impacts their work. The left side of the chart indicates the ratings of the respondents from a scale of -3 to 3. The first box plot in graph A displays whether respondents feel WCA makes their work more difficult/easier, with a median of 1, IQR of 2, mean of 0.78, and standard deviation of 1.45. The second box plot displays whether respondents feel WCA makes their work of the worst/best quality, with a median of 1, IQR of 2, mean of 0.66, and standard deviation of 1.25. The third box plot displays whether respondents feel WCA makes their work slower/faster, with a median of 1, IQR of 2, mean of 0.58, and standard deviation of 1.48. The three box plots look identical. The second graph (B) consists of a bar chart on the top and a box plot on the bottom to display the number of respondents and the self-efficacy ratings. Depicted in the bar chart are the self-efficacy ratings, which include five categories; 48 respondents reported a self-efficacy rating of 1, 104 respondents reported a self-efficacy rating of 2, 281 respondents reported a self-efficacy rating of 3, 191 respondents reported a self-efficacy rating of 4, and 36 respondents reported a self-efficacy rating of 5. The box plot displays the median at 3.25 with an IQR of 1.0, a mean of 3.2, and a standard deviation of 0.93. The third graph (C) also consists of a bar on the top and box plot on the bottom to display the number of respondents and the quality ratings. Depicted in the bar chart are the quality ratings, which include 5 categories: 12 respondents reported a quality of ``Very Poor,'' 82 respondents reported a quality of ``Poor,'' 300 respondents reported a quality of ``Acceptable,'' 184 respondents reported a quality of ``Good,'' and 26 respondents reported a quality of ``Very Good.'' The box plot displays the median at 3.2 with an IQR of 1.0, a mean of 3.2, and a standard deviation of 0.93.}
    \label{fig:survey-productivity}
\end{figure*}

Figure~\ref{fig:survey-productivity}a shows detailed distributions of self-reported ratings of effort, quality of work, and speed (survey question 14). These items were rated on 7-point semantic differential scales, centered on 0. In Figure~\ref{fig:survey-productivity}b, we show the distribution of self-efficacy scores. In Figure~\ref{fig:survey-productivity}c, we show the distribution of quality scores.

\section{Views on code authorship across co-creative scenarios}

\begin{figure*}[ht]
    \centering
    \begin{subfigure}{0.48\textwidth}
        \includegraphics[width=\linewidth]{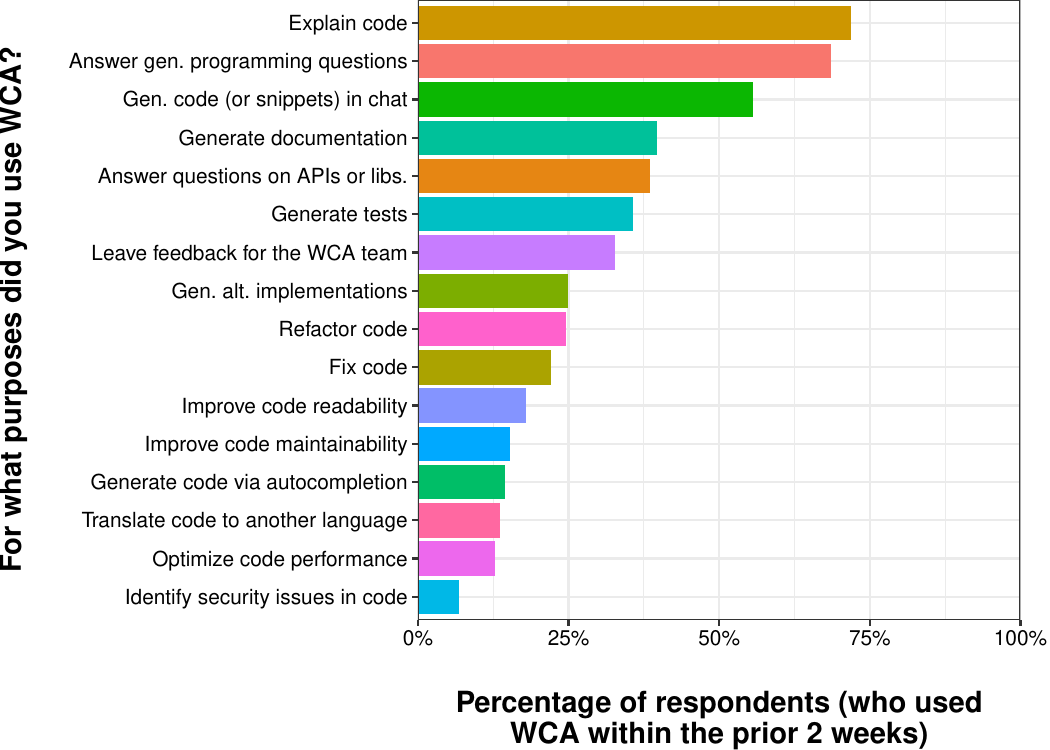}
        \caption{Purposes of use of WCA for the prior two weeks.}
        \label{fig:survey-purposes}
        \Description{The figure displays the distribution of purposes of use of WCA as a horizontal bar graph. There are sixteen categories and the bars indicate the percentage of respondents who indicated using WCA for that purpose. The first purpose is ``Explain code'' (71.9\%). The second purpose is ``Answer general programming questions'' (68.5\%). The third purpose is ``Generate code (or snippets) in chat'' (55.6\%). The fourth purpose is ``Generate documentation'' (39.7\%). The fifth purpose is ``Answer questions with APIs or libraries'' (38.6\%). The sixth purpose is ``Generate tests'' (35.7\%). The seventh purpose is ``Leave feedback for the WCA team'' (32.8\%). The eighth purpose is ``Generate alternate implementations'' (24.9\%). The ninth purpose is ``Refactor code'' (24.6\%). The tenth purpose is ``Fix code'' (22.1\%). The eleventh purpose is ``Improve code readability'' (17.9\%). The twelfth purpose is ``Improve code maintainability'' (15.2\%). The thirteenth purpose is ``Generate code via autocompletion'' (14.4\%). The fourteenth purpose is ``Translate code to another language'' (13.6\%). The fifteenth purpose is ``Optimize code performance'' (12.9\%). The sixteenth purpose is ``Identify security issues in code'' (6.7\%).}
    \end{subfigure}
    \begin{subfigure}{0.48\textwidth}
        \includegraphics[width=\linewidth]{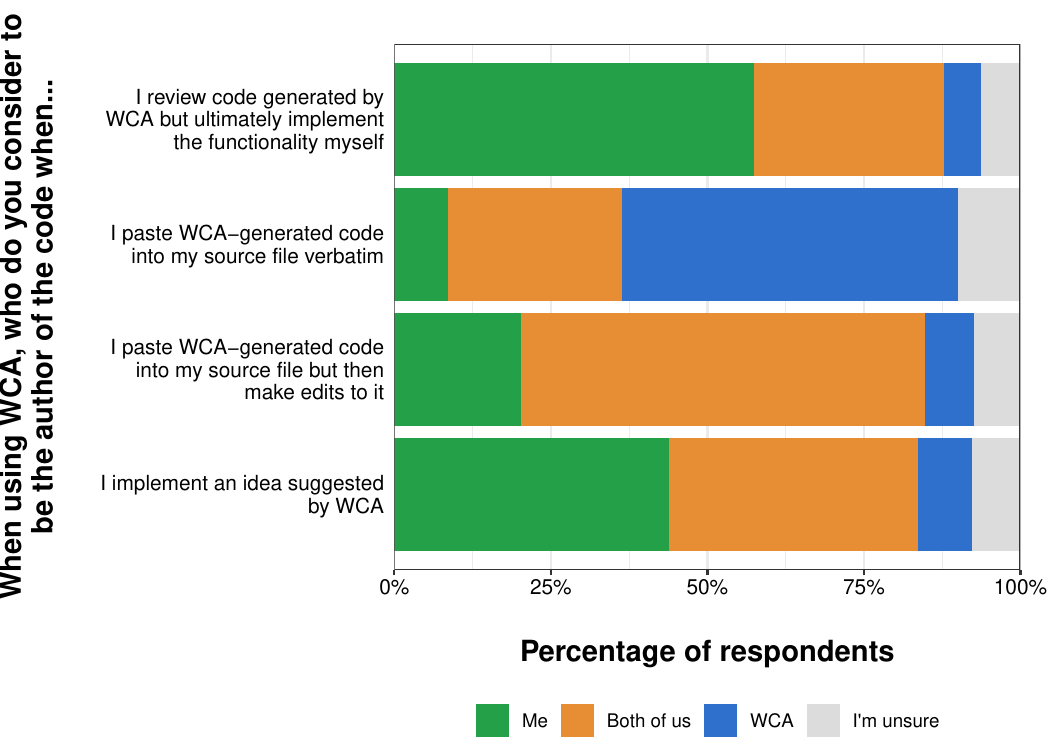}
        \caption{Views on code authorship across different co-creative scenarios.}
        \label{fig:authorship}
        \Description{The figure displays the views of the respondents on code authorship across different co-creative scenarios. It breaks down the percent of respondents who believe ``Me'' to be the author, ``Both of us'' to be the author, ``WCA'' to be the author, and ``Unsure'' of who the author is for each of the four scenarios. The first scenario is ``I review code generated by WCA but ultimately implement the functionality myself'' and the largest percentage group is ``Me'' with 57.5\% of respondents. The second scenario is ``I paste WCA-generated code into my source file verbatim'' and the largest percentage group is ``WCA'' with 53.7\% of respondents. The third scenario is ``I paste WCA-generated code into my source file but then make edits to it'' and the largest percentage group is ``Both of us'' with 64.4\% of respondents. The final scenario is ``I implement an idea suggested by WCA'' and the largest percentage groups are ``Me'' with 43.9\% of respondents and ``Both of us'' with 39.8\% of respondents.}
    \end{subfigure}
    \caption{Distributions of purposes of use of WCA and views on code authorship.}
\end{figure*}

Figure~\ref{fig:authorship} shows the distribution of respondents' views on code authorship across different co-creative scenarios (survey question 16a). Although the major trends align with our intuitions (e.g. when a party authors code, they are an author), we note that across all scenarios, some proportion of respondents felt a joint ownership with WCA.


\end{document}